\documentclass[prd,aps,amssymb,nofootinbib,showpacs,preprint]{revtex4}
\usepackage[russian,english]{babel}
\usepackage[T2A]{fontenc}
\usepackage{amsmath,amsfonts}
\usepackage{textcomp,color}
\usepackage{graphics,graphicx}
\usepackage{hyperref}
\usepackage{overpic}
\begin{document}
\title{Sub-Planckian Scale and Limits for f(R) Models}

\author{\firstname{Polina M.}~\surname{Petriakova}}\email{petriakovapolina@gmail.com} \affiliation{National Research Nuclear University MEPhI (Moscow Engineering Physics Institute),\\ 115409, Kashirskoe shosse 31, Moscow, Russia}
\author{\firstname{Arkady A.}~\surname{Popov}} \email{arkady\_popov@mail.ru}\affiliation{N.~I.~Lobachevsky Institute of Mathematics and Mechanics, Kazan  Federal  University, \\ 420008, Kremlevskaya  street  18,  Kazan,  Russia}
\author{\firstname{Sergey G.}~\surname{Rubin}}\email{sergeirubin@list.ru} \affiliation{National Research Nuclear University MEPhI (Moscow Engineering Physics Institute),\\ 115409, Kashirskoe shosse 31, Moscow, Russia}
\affiliation{N.~I.~Lobachevsky Institute of Mathematics and Mechanics, Kazan  Federal  University, \\ 420008,
Kremlevskaya  street  18,  Kazan,  Russia}

\begin{abstract}
We study the Universe evolution starting from the sub{-}Planckian scale to present times.  The requirement for an exponential expansion of the space with the observed metric as a final stage leads to significant restrictions on the parameter values of a $f(R)${--}function. An initial metric of the Universe is supposed to be maximally symmetric with the positive curvature.
\end{abstract}
\maketitle
\section{Introduction}

It is generally believed that our Universe originated from Planck energies and evolved by expanding and cooling to its present state. The~initial stage of quick expansion starting from the sub-Planckian energy density seems inevitable. {We regard the sub-Planck scale as the highest energy scale in which classical behavior can dominate. The~Planck scale is characterized by complete dominance of quantum fluctuations.} The~spontaneous creation of an inflationary universe is described in detail, for~example, in Reference
~\cite{2004JHEP...09..060F}.
At the same time, the~effects associated with the quantization of gravity may be responsible for model parameter alternation if the energy scale is large enough. Additionally, the~gravity quantization leads to a nonlinear geometric extension of the Einstein--Hilbert~action. 
The first and most successful formulation of the inflationary model, the~Starobinsky model~\cite{1980PhLB...91...99S}, considers nonlinear geometric terms belonging to the $f(R)$ class of theories. Gravity with higher derivatives is widely used in modern research~\cite{Nojiri_2017}, despite the internal problems inherent in this approach~\cite{1988PhLB..214..515B,2015arXiv150602210W}. Attempts were  made to avoid Ostrogradsky instabilities~\cite{2017PhRvD..96d4035P}, and~$f(R)$-gravity was one of the simplest extensions of Einstein--Hilbert gravity free from Ostrogradsky instability.
A necessary element of such models is the fitting of the model parameters to reconstruct the Einstein--Hilbert gravity at low energies~\cite{Fabris:2019ecx}. For~example, in~Reference~\cite{2020CQGra..37w5005O}, the~authors reconstructed the form of the function $f(R)$ using the boundary conditions imposed on the scale factor so that it satisfied the observations in the early and late stages of the evolution of the universe. A~~variety of ways to study the nonlinear multidimensional gravity was discussed in Reference~\cite{Rubin:2020kiy}.

A wide variety of functions $ f (R) $ are presented in the literature. As explicit examples, it is worth citing a couple of functions that relate to a wide range of $f(R)$ functions. The specific model of $f(R)$ gravity
\begin{equation}
    f(R) = R - 2 \Lambda \left(1-\text{e}^{-\frac{\beta R}{2 \Lambda}}\right)\left[1-\frac{\gamma R}{2\Lambda}\log \frac{R}{4\Lambda}\right]
\end{equation}
is considered in Reference~\cite{2018arXiv180702163O}. This model unifies the early time inflationary era and the~late time acceleration of the universe expansion. The~~authors investigated the viability of~the~model and obtained corresponding constraints on free~parameters.

The Tsujikawa model~\cite{2008PhRvD..77b3507T} is in agreement with the cosmological observations~\cite{2019PDU....2600375C} but is slightly different from the $\Lambda$CDM 
model predictions. The~~function $f(R)$ chosen there has the following form:
\begin{equation}
f(R)=R - \lambda R_c \tanh{\frac{R}{R_c}}\, , \quad R_c\, ,\lambda > 0 \, . \end{equation}

Other attempts were undertaken to describe the whole period of evolution of the universe with the multiparametric $f(R)$ function; see~ References \cite{2007JETPL..86..157S,Cognola:2007zu,2020PDU....2900602N,2020PhRvD.101d4009O,2020arXiv201201312O,2020arXiv201200586O,Lyakhova:2018zsr}.

The uncertainty in the parameter values is one of the common questions for such models. To~determine or at least to limit them, the~authors used cosmological and astrophysical observational data, laboratory and solar system tests~\cite{2008PhRvD..77b3507T,Iorio_2019}, binary pulsars, and GW
observations (see~ References \cite{Nojiri:2007as,2009PhRvL.102n1301C,2015CQGra..32x3001B,2015PhR...568....1J,2016RPPh...79d6902K,2012MNRAS.423.3328F,2016A&A...594A..14P,2013LRR....16....9Y} and references therein).

In this article, we discuss the restrictions on the parameters of the following~models based on the known behavior of the scale factor starting from the sub-Planck scale:
\begin{enumerate}
\item  $f(R) = R - 2\Lambda$ ,
\item  $f(R) = a_2 R^2 + R + a_0$ ,
\item  $f(R) = a_3 R^3 + a_2 R^2 + R + a_0$ ,
\item  $f(R) = a_4 R^4 + a_3 R^3 + a_2 R^2 + R + a_0 $,
\end{enumerate}
In~the last three cases, we neglect the cosmological constant compared to the energies we deal with and use $a_0=0$.

We assume that quantum fluctuations nucleate compact Planck-sized manifolds. {Here, we rely on the quantum field theory, where a quantum transition is usually suppressed exponentially by a volume of  nucleated systems.} As the spatial part of the considered four-dimensional metric, we choose the metric of the three-dimensional sphere as the simplest representative:
\begin{equation}\label{ds4}
    ds^2= dt^2 - \text{e}^{2 \alpha(t)}\Bigl(dx^2 + \sin^2{x} \,  dy^2 + \sin^2{x} \, \sin^2{y} \,  dz^2\Bigr)
\end{equation}
 {Other metrics are also nucleated on equal footing, and we plan to study some of them (compact hyperbolic and torus metrics) in the future.}

Constraints on the parameters of the considered models of $ f (R) $ gravity, under~which exponential growth of the scale factor is possible, are investigated.
It is also necessary to determine the conditions under which the exponential growth of the scale factor is replaced by the observed stage of slow expansion. The~~parameters of the model are also limited by the condition that the current size of space must exceed the visible size of the~universe.

{During our study, we kept in mind the following issues:} 

\begin{enumerate}

\item [--] the requirement of  model stability, i.e.,~$f'(R)>0$ and $f''(R)>0$;

\item [--] the quick growth of the space size. It must exceed the size of the visible Universe, $\sim 10^{28}$ cm;

\item [--]extremely small space expansion at the present time.

\end{enumerate}

These requirements are in addition to those usually imposed on the models by the observations at low energies, in~particular, inside the solar~system.

\section{Basic~Equations}

Consider the theory described by  action:
\begin{equation}\label{act0}
S[g_{\mu \nu}]=\frac{m_{Pl}^{2}}{2} \int d^4 x \sqrt{|g|}\, f(R) \, .
\end{equation}
The corresponding extended field equations are as follows:
\begin{equation}\label{eqf(R)}
f_{R} R_{\mu \nu} - \frac{1}{2} \, f g_{\mu \nu} + \Bigl[ \nabla_{\mu} \nabla_{\nu} - g_{\mu \nu} \Box \Bigr]f _{R} =0 \, , \quad \Box \equiv g^{\mu \nu} \nabla_{\mu} \nabla_{\nu}\, , \quad f_R = d f/ d R \, .
\end{equation}
This system of equations coincides with Einstein's field equations for $f(R)=R$.
Throughout this paper, we use the conventions for the curvature tensor $R_{\mu\nu \alpha}^{\beta}=\partial_{\alpha}\Gamma_{\mu \nu}^{\beta}-\partial_{\nu}\Gamma_{\mu \alpha}^{\beta}+\Gamma_{\sigma \alpha}^{\beta}\Gamma_{\nu\mu}^{\sigma}-\Gamma_{\sigma \nu}^{\beta}\Gamma_{\mu \alpha}^{\sigma}$ and the Ricci tensor is defined as $R_{\mu \nu}=R^{\alpha}_{\mu \alpha \nu } \, $.

Let us suppose that the action and metric have the forms 
 \eqref{act0} and \eqref{ds4} consequently.
In~this case, the~ nontrivial Equation \eqref{eqf(R)} acquirse the following form:
\begin{eqnarray}\label{eqsyst1}
&& 6\dot{\alpha}\dot{R}f_{RR} -6 \Bigl(\ddot{\alpha}+\dot{\alpha}^2 \Bigr)f_R  +f(R) = 0 \, , \\
&& 2\dot{R}^2 f_{RRR} +2\Bigl(\ddot{R} +2\dot{\alpha} \dot{R} \Bigr)f_{RR} 
-\Bigl(2\ddot{\alpha} +6\dot{\alpha}^2 +4\text{e}^{-2\alpha} \Bigr)f_R + f(R) = 0 \, ,\label{eqsyst2}
\end{eqnarray}
{where Equation \eqref{eqsyst1} correspond to the (tt)-component and~Equation \eqref{eqsyst2} corresponds to the coinciding components (xx) = (yy)=(zz) of system \eqref{eqf(R)}.} 
The definition of the Ricci scalar for metric \eqref{ds4} is
\begin{equation}\label{R}
    R = 12\dot{\alpha}^2 + 6\ddot{\alpha} + 6\text{e}^{-2\alpha} .
\end{equation}
{Substituting $\ddot{\alpha}$ from \eqref{R} into the Equation \eqref{eqsyst1}, we obtain an equation that does not contain the second derivatives of the functions $\alpha$ and $R$:
\begin{equation}\label{restr}
6\dot{\alpha}\dot{R}f_{RR} + \Bigl(6\dot{\alpha}^2 +6\text{e}^{-2\alpha} -R \Bigr) f_R  +f(R) = 0 \, ,
\end{equation}
There are three Equations \eqref{eqsyst1}--\eqref{R} with respect to the unknown functions $\alpha(t)$ and $R(t)$, but~only two of them are independent. It is technically easier to solve Equations \eqref{eqsyst2} and~\eqref{R}. Equation \eqref{restr} plays the role of a restriction to the solutions of  second-order differential Equations \eqref{eqsyst2} and \eqref{R}. This equation was used twofold. Firstly, this equation should be the identity when the solution of systems  \eqref{eqsyst2} and \eqref{R} are substituted. Secondly, applied at $t=0$, it was used to fix one of the initial variables. 
}

We look for those solutions to this system of equations that have ``correct" asymptotic behavior. The~~latter are those that could describe our universe at present time. The~space size should be not smaller than the size of the universe. Therefore, space should expand extremely quickly, at~least in the beginning. The~~asymptotic value of the Hubble parameter should not be bigger than the observable one. Due to its smallness, compared to the sub-Planckian energies, we use $ H \xrightarrow{t\rightarrow\infty} 0$. That means $\alpha(t) \xrightarrow{t\rightarrow\infty} \text{const}$ and the asymptotic value of the Ricci scalar $R(t) \xrightarrow{t\rightarrow\infty} 0$. 
We also assume that $\alpha(t \rightarrow \infty) > 140,$ where the value $\text{e}^{140}m^{-1}_{Pl}$ corresponds to the horizon scale $10^{28}$cm at present~time.

Knowledge of the asymptotic behavior facilitates the analysis. We sought for the solutions with asymptote $ \alpha(t) \xrightarrow{t\rightarrow\infty} Ht$. Therefore, $R(t) \xrightarrow{t\rightarrow\infty} 12H^2 +6\text{e}^{-2Ht}$ and $\dot{R}(t)\xrightarrow{t\rightarrow\infty} -12H\text{e}^{-2Ht}$. A~t~the end of the asymptotic regime, $R(t=\infty)=R_c=\text{const}$ and $\dot{R}(t=\infty)=0.$
In this case ($R=\text{const}$), the trace of system \eqref{eqf(R)} leads to the algebraic equation
\begin{equation}\label{eqf(Rc)}
f _{R} ( R_{c} ) R_{c} - 2 \,f( R_{c} ) =0 \, . 
\end{equation}
Several solutions of this equation could take place for specific values of the physical parameters of function $f(R)$. The Ricci scalar averaged over large scale is negligibly small at present time. Therefore, our aim is the asymptotic solution $R_c=R_{Universe}\simeq 0$.

Let us fix the initial conditions for systems \eqref{eqsyst2} and \eqref{R} 
\begin{equation}\label{const}
 \alpha(0) = \alpha_0 \, , \quad \dot{\alpha}(0) = \alpha_1 \, , 
 \quad \dot{R}(0) = R_1 \, .
\end{equation}
Restriction \eqref{restr} is used to fix the initial value of the curvature $ R(0) = R_0$.

We are interested in the dynamics of the maximally symmetric manifold starting from the sub{-}Planck scale. Therefore, the natural choice of the initial conditions is
\begin{equation}\label{c1}
\alpha_0 \sim \ln{H_{\text{sub-Planck}}^{-1}} \,\, , \quad  \alpha_1\, \sim H_{\text{sub-Planck}} \,\, , \quad H_{\text{sub-Planck}}\, \lesssim \, m_{Pl} \, .
\end{equation}
Further, we work in the Planck units, $m_{Pl} =1$.

The sections below describe the rate of space growth for several forms of the $f(R)$ function depending on the initial data and physical~parameters.

\section{The Dependence of the Universe Expansion on the Lagrangian Parameters}
\unskip

\subsection{R---Gravity}

In the case $f(R)=R-2\Lambda$, we have a well-known solution:
\begin{equation}
\text{e}^{2 \alpha} = \frac{3}{\Lambda} \cosh^2 \left( \sqrt{\frac{\Lambda}{3}}\, t \right), \quad \Lambda > 0 \, .
\end{equation}
The observations indicate~\cite{2020A&A...641A...6P} that the parameter $\Lambda\sim 10^{-122}$ in the Planck units. Therefore, the~initial size of the manifold nucleated is of the order $10^{61}$. The~~nucleation probability of such a huge volume due to the quantum effects is negligible. Therefore, this model does not satisfy the considered~assumptions.

\subsection{R$^2$---Gravity}\label{Sec3.2}

For a well-studied model
\begin{equation}\label{fR2}
f(R)=a_2 R^2 + R, \,
\end{equation}
the asymptote of the curvature is zero $(R_c =0)$,  which is the solution of Equation \eqref{eqf(Rc)}. That means that $\alpha(t) \xrightarrow{t\rightarrow\infty} \text{const}$. The~~question is what is the size of the finite space? 

\begin{figure}[ht!]
 \begin{center}
 \includegraphics[width=0.36\textwidth]{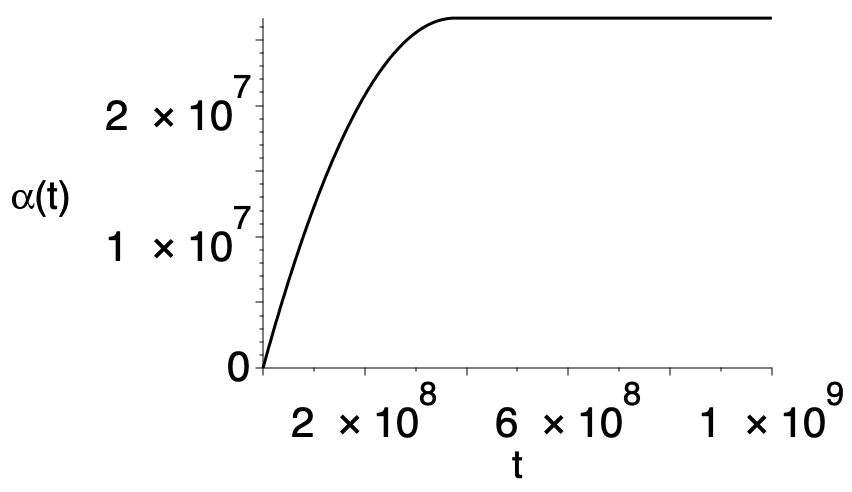}
 \hspace{0.5cm}
 \includegraphics[width=0.36\textwidth]{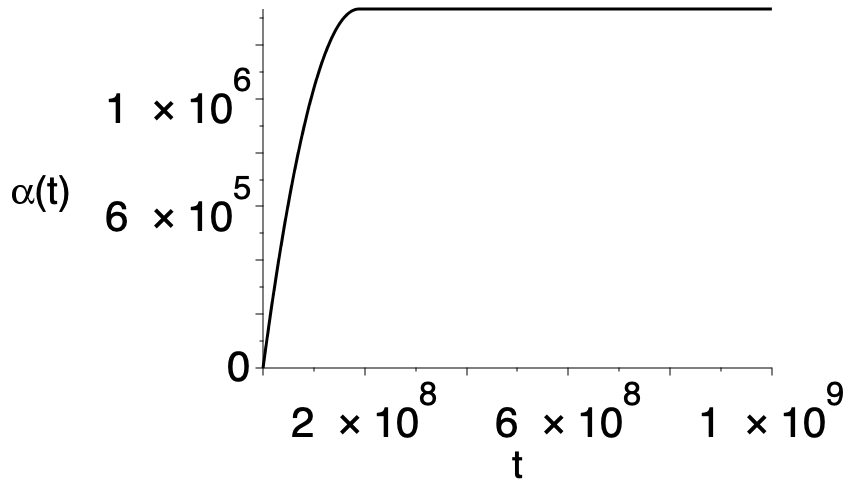}
 \end{center}
\vskip-5mm \caption{ The solution of the system with parameters $a_2 \simeq 10^{9}$ and the initial conditions $ \alpha_0 = 2.3 $, $ \alpha_1 = 0.1 $, $R_0 = 0.24 $, and $R_1 = 0$ (left side) and  $ \alpha_0 = 4.6 $, $ \alpha_1 = 0.01 $, $R_0 = 0.002 $, and $R_1 = 0$ (right side).}
\label{Fig:Sub-Planck_R2}
\end{figure}

Let us find numerically the solution of systems  \eqref{eqsyst2} and \eqref{R} starting from the sub{-}Planckian scale, i.e.,~with initial conditions \eqref{c1} and {the value of the parameter $a_2$ chosen according to the Starobinsky model, $a_2=1/6m^2\simeq 10^{9}$, where $ m/m_{Pl} \sim 10^{-5}$ \cite{2011PhLB..700..157G}.
}
 The result is shown in Figure~\ref{Fig:Sub-Planck_R2}. The~space size is of the order $ \sim \exp{(10^{6 \div 7})} m_{Pl}^{-1}$ by the end of the inflationary stage, which exceeds the size of the visible part of the Universe and, hence, does not contradict~observations.

The model predicts the substantial growth of the space size that looks quite evident. Nevertheless, the~model parameter $a_2$ is too {large} to be natural. This means that there are some processes that occur above the inflationary scale that strongly influence the parameter value. It is the nontrivial subject of future~research.

\subsection{R$^3$---Gravity}

Our next choice is the function
\begin{equation}\label{f(R)3}
f(R)= a_3 R^3 + a_2 R^2 + R \, .
\end{equation}
There are three types of asymptotes following from algebraic Equation \eqref{eqf(Rc)}
\begin{eqnarray}\label{f3tq}
R_c \left( a_3 R_c^2 -1 \right)=0 \quad \Rightarrow \quad \{R_c\}_{1} = 0  \, , \,\, \{R_c\}_{2,3} = \pm \frac{1}{\sqrt{\,a_3\,}} \, .
\end{eqnarray}
The first one is realized in our Universe. Our immediate task is to find the solutions to Equations \eqref{eqsyst1} and  \eqref{eqsyst2} that lead to the observed Universe.
The aim is to impose restrictions on the model parameters $a_2$ and $a_3$ by the analysis of the metric dynamic starting from the sub-Planckian scale. It is assumed that nucleated manifolds should expand up to the~observable~size. 

The necessary conditions for the behavior of solutions to equations are listed in the Introduction. In~short, not only should the~criterion be $ R_c = 0 $ but also~the solution should be stable, should grow rapidly from the very beginning, and should strive for a constant at the final stage. 
The numerical solution was found by the Rosenbrock method for the Cauchy problem in the Maple computer mathematics system.

\begin{figure}[th!] 
 \begin{center}
 \includegraphics[width=0.3\textwidth]{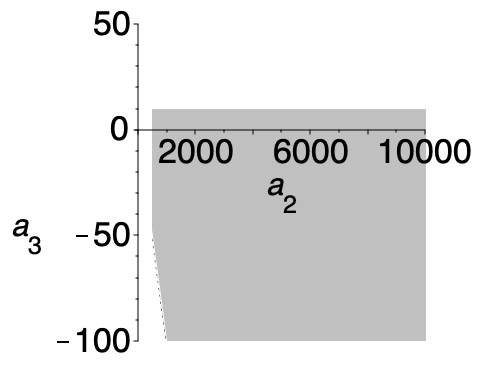} \hspace{0.5cm}
 \includegraphics[width=0.3\textwidth]{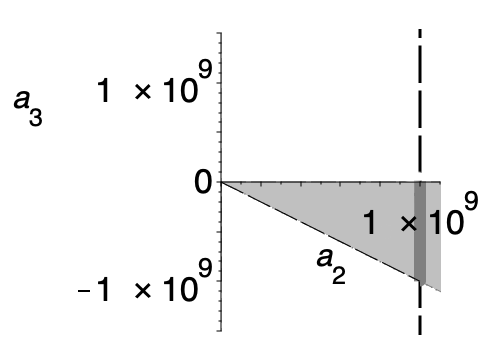}
 \end{center}
\vskip-5mm \caption{The range of values of the parameters $a_2$ and $a_3$ that leads to the space of a size larger than the visible part of the universe. The~initial conditions were chosen as $ \alpha_0 = 2.3 $, $ \alpha_1 = 0.1 $, and $R_1 = 0$. The~intersection of the constraints obtained in our analysis and by the authors in~\cite{2020PhLB..80535453C} occurs in the approximate range of value $a_3 \in [-10^9, 10^{-6}]$.}
\label{ParamPhase}
\end{figure}

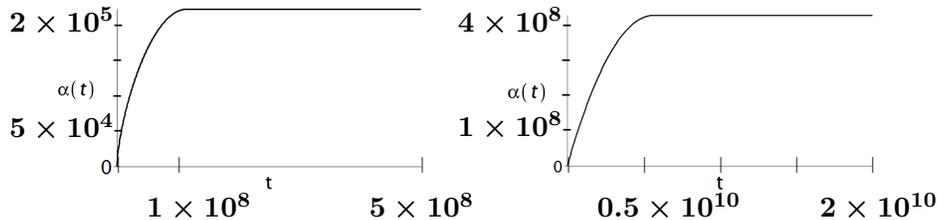
\begin{figure}[th!]
 \begin{center}
\begin{overpic}[width=0.30\textwidth]{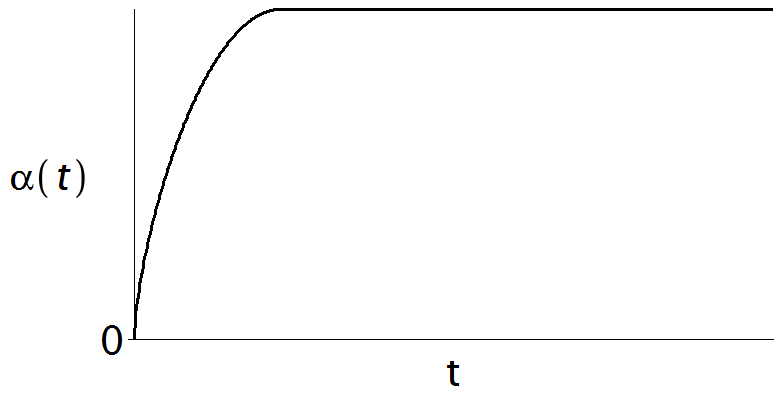}
\put(-12,44){\small{\boldmath $2\times10^5 $}-}
\put(16,34.5){-}
\put(16,25){-}
\put(-12,15.5){\small{\boldmath $5\times10^4 $}-}
\put(16.6,7){\tiny{$|$}}
\put(33,7){\tiny{$|$}}
\put(98.5,7){\tiny{$|$}}
\put(25,-6){\small{\boldmath $1\times10^8 $}}
\put(85,-6){\small{\boldmath $5\times10^8 $}}
\end{overpic}
 \hspace{0.8cm}
\begin{overpic}[width=0.30\textwidth]{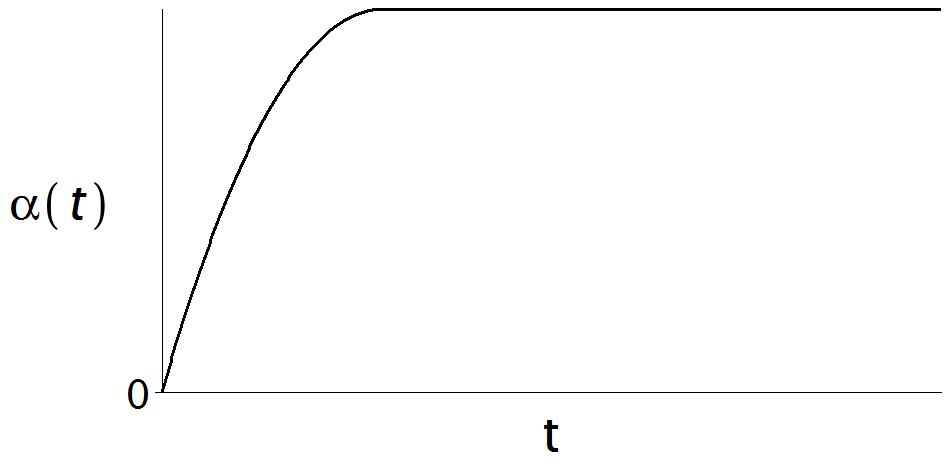}
\put(-12.5,44){\small{\boldmath $4\times10^8 $}-}
\put(-12.5,15.88){\small{\boldmath $1\times10^8 $}-}
\put(15.5,34.63){-}
\put(15.5,25.25){-}
\put(16.5,7){\tiny{$|$}}
\put(37,7){\tiny{$|$}}
\put(57.5,7){\tiny{$|$}}
\put(78,7){\tiny{$|$}}
\put(98.5,7){\tiny{$|$}}
\put(25,-6){\small{\boldmath $0.5\times10^{10} $}}
\put(85,-6){\small{\boldmath $2\times10^{10} $}}
\end{overpic}
 \end{center}
\vskip-5mm \caption{ The solution of the system with parameters $a_3 = -10^{8}$ and $a_2 = 10^{9}$ (left side); $a_3 = -1$ and $a_2 = 10^{9}$ (right side); and the~initial conditions $ \alpha_0 = 2.3 $, $ \alpha_1 = 0.1 $, $R_0 = 0.29$, and $R_1 = 0.01$. }
 \label{Fig:Sub-Planck_R3}
\end{figure}

The numerical analysis leads to the following limits represented in the phase diagram of~Figure~\ref{ParamPhase}.
 The acceptable region obtained here is marked in gray.  Those values of $a_3$ are not acceptable since the other conditions are not satisfied (the solution $R_c=0$ is not stable and/or the space growth is too slow). 
The boundaries are smooth due to a possible variation in the initial conditions in Equation \eqref{c1}. It is assumed that the manifolds are nucleated due to the quantum effects at the sub-Planckian scale so that the probability of large sized manifolds is negligible. The~part of the acceptable parameter region derived in Reference ~\cite{2020PhLB..80535453C}
is marked by a black dashed line (the right panel).
The common area belonging to both restrictions is much less than each of~them.
Appropriate results can be obtained not only for the trivial initial condition $\dot{R}(0)= R_1 = 0$;  see as an example Figure~\ref{Fig:Sub-Planck_R3} with  $\dot{R}(0)= R_1 = 0.01$.

\subsection{R$^4$ -~Gravity}

\begin{figure}[ht!]
 \begin{center}
 \includegraphics[width=0.36\textwidth]{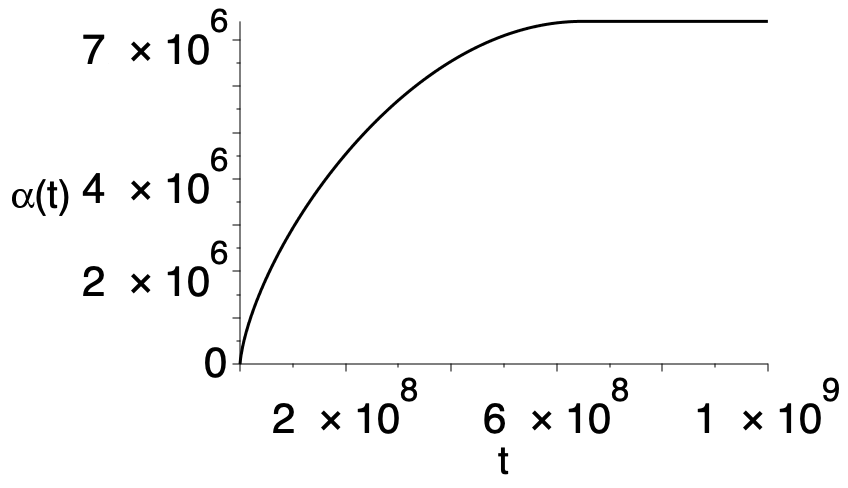}
 \hspace{0.5cm}
 \includegraphics[width=0.36\textwidth]{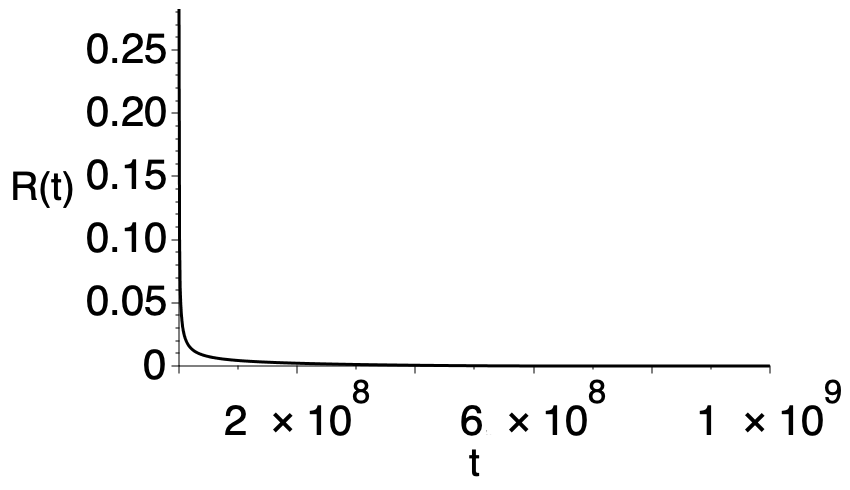}
 \end{center}
\vskip-5mm \caption{ The solution of the system with parameters $a_4 = -10^{2}$, $a_3 = -10^{5}$, and $a_2 = 10^{9}$ and the initial conditions $ \alpha_0 = 2.3 $, $ \alpha_1 = 0.1 $, $R_0 = 0.28$, and $R_1 = 0.01$.}
\label{Fig:Sub-Planck_R4}
\end{figure}

\begin{figure}[ht!]
\begin{center}
\begin{overpic}[width=0.5\textwidth]{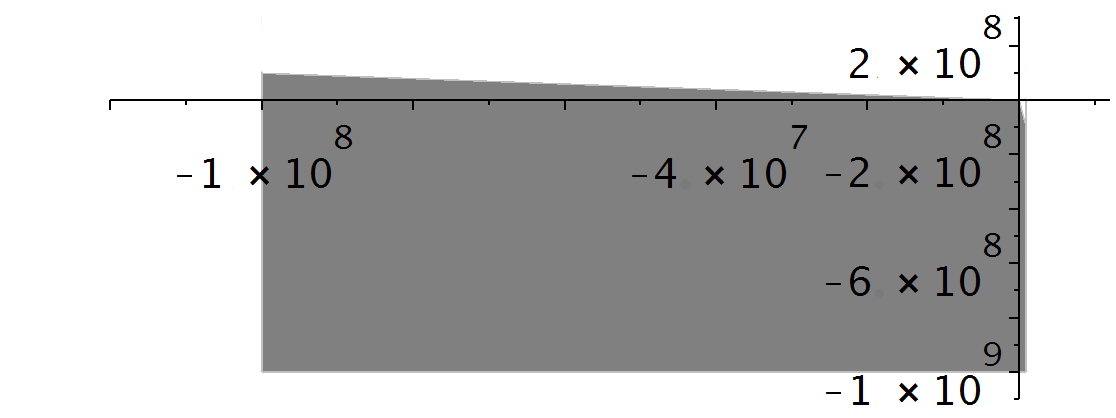}
\put(40,33){\textrm{$ a_3$}} 
\put(93,13){   \textrm{$ a_4$}}
\end{overpic}
\end{center}
\vskip-5mm \caption{The range of the acceptable parameter values $a_4$ and $a_3$. The~~initial conditions are $ \alpha_0 = 2.3 $, $ \alpha_1 = 0.1 $, and $R_1 = 0$.}
\label{ParamPhaseR4}
\end{figure}

As a final example, consider the function
\begin{equation}\label{f(R)_R4}
    f(R) = a_4 R^4 + a_3 R^3 + a_2 R^2 + R \, , 
\end{equation}
with the most realistic estimation of the parameter  $a_2 \sim 10^9$ according to the discussion in \mbox{Section~\ref{Sec3.2}.} A typical behavior of the metric is shown in Figure~\ref{Fig:Sub-Planck_R4}. The~phase diagram for parameters $a_4$ and $a_3$ at fixed $a_2 \sim 10^9$ is presented in Figure~\ref{ParamPhaseR4}. The~space expands sufficiently if the~parameter values $(a_3,a_4)$ belong to the gray area. The~area borders are slightly smoothed if the initial conditions are varied. The~boundary shift is small and does not influence the~conclusion.

\nointerlineskip

\section{Conclusions}

In this paper, we discuss new restrictions imposed on the parameters of some $f(R)$ models of the gravity. These restrictions are the result of studying the Universe evolution at high energies. 
We suppose that our Universe was nucleated with the size of the Planck scale order. It must expand rapidly to reach a size no smaller than that of our Universe at present time. We also choose the 3 dimensional spherical metric from the very beginning as the additional~assumption. 

These suppositions being quite natural lead to new restrictions compared to limits based on the observations {in the Solar system}. For~example,  the~parameter range of $R^3$ gravity is severely tightened if we apply both our restriction and those in the paper of~\cite{2020PhLB..80535453C}. 
In all models discussed here, the~parameter ranges depend on the initial conditions that lead to their slight uncertainties. Nevertheless, these restrictions should be taken into account in the considered models based on gravity with higher derivatives. {It is worth mentioning that the pure Einstein--Hilbert gravity with the $\Lambda$ term is not realized in the framework of our approach.}
\section*{Acknowledgments}
This research was funded by the Ministry of Science and Higher Education of the Russian Federation, Project ``Fundamental properties of elementary particles and cosmology'' N 0723-2020-0041.
The work of A.P.  and S.R. was supported by the Kazan Federal University Strategic Academic Leadership Program. The~work of A.P. was partly funded by the Russian Foundation for Basic Research grant No. 19-02-00496.
The work of A.P was also funded by the development program of the Regional Scientific and Educational Mathematical Center of the Volga Federal District, agreement N~075-02-2020.

The authors declare no conflict of interest.

\end{document}